\documentclass[12pt]{article}
\makeatletter
\renewcommand{\@seccntformat}[1]{%
        {\csname the#1\endcsname}\ \ }

\renewcommand{\section}{%
    \@startsection{section}{1}{\z@}%
    {-3.5ex plus -1ex minus -.2ex}%
    {2.3ex plus.2ex}%
    {\centering\normalsize\bfseries}}

\renewcommand{\subsection}{\@startsection{subsection}{2}{0pt}%
        {-3.25ex plus -1ex minus -.2ex}%
        {1.5ex plus .2ex}%
        {\centering\normalsize\itshape}}

\makeatother

\newcommand{\drawsquare}[2]{\hbox{%
\rule{#2pt}{#1pt}\hskip-#2pt%  left vertical
\rule{#1pt}{#2pt}\hskip-#1pt%  lower horizontal
\rule[#1pt]{#1pt}{#2pt}}\rule[#1pt]{#2pt}{#2pt}\hskip-#2pt%  upper horizontal
\rule{#2pt}{#1pt}}% right vertical

\newcommand{\Delamb}{\drawsquare{7}{0.6}}

\newcommand{\tr}{\mathop{\hbox{\rm tr}}}

\def\lesssim{\mathrel{\mathpalette\vereq<}}
\def\gtrsim{\mathrel{\mathpalette\vereq>}}
\makeatletter
\def\vereq#1#2{\lower3pt\vbox{\baselineskip1.5pt \lineskip1.5pt
\ialign{$\m@th#1\hfill##\hfil$\crcr#2\crcr\sim\crcr}}}
\makeatother

\renewcommand{\thefootnote}{\fnsymbol{footnote}}
\begin{document}

\begin{titlepage}
\begin{center}
\today     \hfill    CERN-TH/98-337\\
~{} \hfill LBNL-42419 \\
~{} \hfill UCB-PTH-98/50\\
~{} \hfill UMD-PP99-037\\
~{} \hfill hep-ph/9810442\\

%\vskip .25in
\vskip .1in

{\large \bf Gaugino Mass without Singlets}%
\footnote{This work was supported in part by the U.S.
Department of Energy under Contracts DE-AC03-76SF00098, in part by the
National Science Foundation under grants PHY-95-14797
and PHY-98-02551, and by the Alfred P. Sloan Foundation.}

\vskip 0.3in

Gian F. Giudice$^\dagger$, Markus A. Luty$^\ddagger$,
Hitoshi Murayama$^{**}$, and Riccardo Rattazzi$^\dagger$

\vskip 0.05in

{\em $^\dagger$Theory Division, CERN\\Geneva, Switzerland}

\vskip 0.05in

{\em $^\ddagger$Department of Physics, University of Maryland\\
College Park, Maryland 20742, USA}

\vskip 0.05in

{\em $^{**}$Department of Physics,
University of California\\
Berkeley, California 94720, USA}

\vskip 0.05in

\end{center}

\vskip .1in

\begin{abstract}
\noindent
In models with dynamical supersymmetry breaking in the hidden
sector, the gaugino masses in the observable sector have been
believed to be extremely suppressed (below 1 keV), unless there is
a gauge singlet in the hidden sector with specific couplings to the
observable sector gauge multiplets.
We point out that there is a pure supergravity contribution to
gaugino masses at the quantum level arising from the
superconformal anomaly.
% Our basic point is that
% one must rely on the full 1PI action in a supergravity
% breaking background, rather than on a local effective lagrangian.
Our results are valid to all orders in perturbation theory and are related to
the `exact' beta functions for soft terms.
There is also an anomaly contribution to the $A$ terms proportional to
the beta function of the corresponding Yukawa coupling.
The gaugino masses are proportional to the corresponding gauge beta
functions, and so do not satisfy the usual GUT relations.
%We point out that
%supersymmetric CP problem is substantially ameliorated within this
%scheme.
%
% Most of the discussion applies also to hidden sector models
% without $F$-term; however we point out that flavor problem is not
% solved in these models, contrary to a claim in the literature.
\end{abstract}

\end{titlepage}

\newpage
\renewcommand{\thefootnote}{\arabic{footnote}}%  use numbers
\setcounter{footnote}{0}

\section{Introduction}
% Supersymmetry (SUSY) breaking communicated by supergravity (SUGRA)
% interactions is arguably one of the most attractive scenario
% for the realization of SUSY in nature.
% In models of this type, SUSY is broken in a hidden sector
% and gravitational-strength interactions communicate SUSY breaking
% to the observable sector.
% The main advantage of this scenario lies in its theoretical appeal:
% the key ingredients are either present of necessity ({\it e.g.} SUGRA)
% or very well-motivated
% ({\it e.g.} hidden sectors are generically present in string theories).
% The main disadvantage of this scenario is that at present there is no
% convincing explanation for the degeneracy of squark masses required
% to avoid large flavor-changing neutral current effects.
% In the context of string theory and SUGRA models with singlets,
% there are also cosmological problems related to
% the existence of uncharged fields with almost flat potentials
% and interactions suppressed by powers of the Planck scale.

Supersymmetry (SUSY) is arguably the most attractive mechanism to
stabilize the hierarchy between the fundamental scale ({\it e.g.}~the
Planck scale $M_* \sim 10^{18}$~GeV) and the electroweak scale
($M_{W} \sim 100$~GeV).
However, superpartners of the standard-model particles have not been
observed up to energies of order $M_W$, so SUSY must be broken at or
above the weak scale.
The phenomenology of SUSY depends crucially on the mechanism of SUSY
breaking and the way that SUSY breaking is
communicated to the observable sector.

Communication of SUSY-breaking effects by supergravity (SUGRA)
interactions is in some ways the most attractive scenario.
In models of this type, SUSY is broken in a hidden sector
and gravitational-strength interactions communicate SUSY breaking
to the observable sector.
The main advantage of this scenario lies in its theoretical appeal:
the key ingredients are either present of necessity ({\it e.g.} SUGRA)
or very well-motivated
({\it e.g.} hidden sectors are generically present in string theories).
The main disadvantage of this scenario is that at present there is no
convincing explanation for the degeneracy of squark masses required
to avoid large flavor-changing neutral current effects.
In the context of string theory and SUGRA models with singlets,
there are also cosmological problems related to
the existence of uncharged fields with almost flat potentials
and interactions suppressed by powers of the Planck scale.

In order to explain the origin of the SUSY breaking scale (and hence
the weak scale) the most attractive scenario is that SUSY is broken
dynamically~\cite{ADS1,ADS2,ADS3}.
In recent years, it has been found that this occurs in many
asymptotically-free supersymmetric gauge theories.
In these models, dimensional transmutation generates the hierarchy
between the SUSY breaking scale $\mu_{\rm SUSY}$ and the Planck scale,
and the SUSY-breaking masses are of order $\mu_{\rm SUSY}^2 / M_*$.
%
%  ({\it e.g.} scalar masses) appear
% in the supersymmetric standard model at $\sim \mu_{SUSY}^{2}/M_{Pl}$.
%
%
% In order to have a model with vanishing cosmological constant and
% broken SUSY, the theory must break SUSY in the flat-space limit
% \cite{Weinberg}.%
% %
% \footnote{For example SUSY is broken by gaugino condensation in
% super Yang--Mills theory in the context of SUGRA, but there is a
% negative cosmological constant.}
% %
The most important challenge of constructing phenomenologically viable
models of dynamical SUSY breaking in the hidden sector is generating
sufficiently large gaugino masses \cite{ADS3,DM}.
In models without gauge singlets in
the hidden sector, the gaugino mass is conventionally believed
to be extremely
suppressed, at most of order $\mu_{\rm SUSY}^{3}/M_{*}^{2} \simeq 1$~keV.
% and it is often even more suppressed.
There have been a variety of
solutions discussed in the literature \cite{BKN,IY1,Nelson}, all of
which involve gauge singlets with SUSY-breaking VEV's, and require
more or less complicated model-building.
It is not at all clear whether any of these solutions can work in the
context of string theory, where one singlet field, the dilaton, couples
to all gauge kinetic terms.
Obtaining realistic gaugino masses in string theory therefore appears
to require a large $F$ component for the dilaton (in addition to the
usual dilaton stabilization problem), which does not occur in
conventional mechanisms for (locally) stabilizing the
dilaton.\footnote{One could still use $F$-component of moduli fields
  which appear in the gauge kinetic function at the one-loop level.
  Here again the stabilization is an issue, and the cosmological problem is
there as well.}
More generally,
the presence of gauge singlet fields also causes a variety of concerns,
such as cosmological problem
\cite{Coughlan,Roulet,BKN} or destabilization of hierarchy \cite{BP}.

In this paper, we point out a completely
model-independent contribution to the
gaugino mass whose origin can be traced to the conformal anomaly.
% Contrary to earlier claims in the literature \cite{ADS3,DM,BKN},
This contribution is always present even if there are no gauge singlet
fields that generate the gaugino masses at the tree-level.
Therefore, no model-building gymnastics is necessary to generate
gaugino masses at order $1/M_{*}$.
This contribution to the gaugino mass is given {\it exactly}\/
(to all orders in perturbation theory) by
\begin{equation}
m_\lambda = \frac{\beta(g^2)}{2 g^2} m_{3/2},
\label{prima}
\end{equation}
where $\beta(g^2) = d g^2 / d \ln\mu$ is the gauge beta function.
In models without singlets (or models in which the singlets do not
couple to the gauge fields in the required way), Eq.~(\ref{prima})
gives the leading effect in the gaugino mass.
This has interesting phenomenological consequences.
First, the gaugino mass ratios are
given by ratios of beta functions, a very different result from the
usual `unified' relation.
Other aspects of the phenomenology depend crucially on the scalar masses.
The simplest assumption is that the scalar masses are of order
$m_{3/2}$, which is much larger than the gaugino masses in
Eq.~(\ref{prima}).
This scenario unfortunately suffers from quite severe fine-tuning
required for electroweak symmetry breaking, but has a predictive and
interesting phenomenology that we will discuss below.
An alternative possibility is that the scalar masses are naturally
suppressed compared to $m_{3/2}$.
For example, this occurs in models with Heisenberg
symmetry \cite{Heisenberg}, {\it i.e.}\/, models of
`no-scale' type \cite{LN}.

In complete analogy to Eq.~(1), we also show that the $A$-terms arise
proportionally to the $\beta$-function of the corresponding Yukawa coupling.

Contributions to gaugino masses that are proportional to the
corresponding $\beta$-functions have been previously found in
the string-based models
of Ref.~\cite{ibanez}, using the results in Ref.~\cite{zwirner}.
 However, those contributions depend on the moduli and
therefore, unlike Eq.~(1), their normalization is not purely
fixed by the gravitino mass. We emphasize that the contribution considered here
exists in any model, and becomes the dominant one in particular
classes.
%In spite of these important theoretical distinctions, the
%phenomenology
%of some of the models of Ref.~\cite{ibanez} may be  similar
%to the one we consider here.

This paper is organized as follows.  In Section 2, we review the
important features of dynamical SUSY breaking in the hidden sector,
and comment on previous work on gaugino masses.  Section 3 contains
our main results.
We derive formulae for gaugino
masses and other ${\cal O}(m_{3/2})$ SUSY-breaking parameters to all
orders in perturbation theory in models without gauge singlets
in the hidden sector.
In Section 4 and 5, we consider phenomenology and the `$\mu$ problem.'
Section 6 contains our conclusions.

%-----------------------------------------------------------------------
%\section{Hidden Sectors without Planck-scale VEV's or Singlets}
\section{Dynamical SUSY Breaking in the Hidden Sector without Singlets}
\label{sec:generality}
%-----------------------------------------------------------------------
In this section, we review the main features of SUGRA models with
dynamical SUSY breaking in the hidden sector and no singlets.
Our primary motivation for dynamical SUSY breaking
is that it is the simplest mechanism for generating
the SUSY breaking scale, and hence explaining (rather than simply
stabilizing) the hierarchy between the weak scale and the Planck scale.
(In fact, if we want to have a SUSY breaking scale well below the
Planck scale, and we assume that the K\"ahler potential is `generic',
it can be shown that SUSY must be broken in the flat limit
\cite{Weinberg,Nelson}.)
We consider models without singlets because we will show below
that they are not necessary to obtain large gaugino masses.

% dynamical SUSY breaking
% models do not generally involve singlets with sufficiently large
% SUSY-breaking VEV's to give rise to gaugino masses of the same size
% as scalar masses.
% On the other hand, the main result of the present paper is
% a model-independent ${\cal O}(m_{3/2})$ contribution to the gaugino
% mass that exists even if the model contains no singlets.

We therefore consider a model that breaks SUSY dynamically at a scale
$\mu_{\rm SUSY}$ in the flat
limit $M_* \to \infty$, and couple it to SUGRA.
Since the model has a stable vacuum in the flat limit, we do not expect
any Planck-scale VEV's.%
\footnote{Even for models with
non-renormalizable interactions suppressed by powers of Planck
scale, the expectation values are often much smaller than the Planck
scale.}
This is to be contrasted with the situation in conventional hidden sector
models, in which generally there are
fields with VEV's of order (or larger than)
the Planck scale \cite{AN,BFS,HLW}.
In models without Planck-scale VEV's,
% We therefore consider models that spontaneously break global SUSY
% at a scale $\mu_{\rm SUSY} \sim 10^{11}$~GeV, coupled to SUGRA.
% In order to understand the origin of the SUSY breaking scale, we
% assume that SUSY is broken dynamically.
% If we write the theory in terms of fields that do not have Planck-scale
% expectation values,
the SUGRA scalar potential simplifies drastically.
By keeping the leading ${\cal O}(\mu_{\rm SUSY}^4)$
terms of an expansion in $\mu_{\rm SUSY}/M_*$,
one finds
\begin{equation}
  V = |W_z|^2 - \frac{3}{M_*^2} |W|^2 + \mbox{$D$-terms}
  + {\cal O}(\mu_{\rm SUSY}^5),
\end{equation}
irrespective of the form of the K\"ahler potential as long as it
has a Taylor expansion with canonical kinetic term as its lowest order
term: $K = z^* z + {\cal O}(z^3/M_*)$.
(A linear term is absent if there are no singlets.)
The first term is equivalent to the case of globally supersymmetric theories
and has a finite (positive) value as long as SUSY is broken.
The second term is used to fine-tune the cosmological constant by adding
a constant term %of ${\cal O}(\mu_{SUSY}^2 M_*)$
in the superpotential, related to
the gravitino mass by $\langle W \rangle = m_{3/2} M_*^2$.

The soft terms in the observable sector described by the fields
$\phi$ come from the cross terms in
$(K_i W + W_i)^* K^{-1}_{ij} (K_j W + W_j) = |W_i|^2 + m_{3/2} (\phi_i
W_i + {\rm h.c.}) + {\cal O}(m_{3/2}^2)$ and $-3|W|^2 = -3 m_{3/2} W +{\rm
h.c.}+{\cal O}(m_{3/2}^2)$.
Therefore, ${\cal O}(m_{3/2})$ terms are completely model-independent,
\begin{equation}
  m_{3/2} (\phi_i W_i - 3 W), \label{eq:universal}
\end{equation}
and hence $A=0$, $B=-m_{3/2}$ and $C=-2m_{3/2}$ \cite{JY}.\footnote{It
  is interesting that this particular form of the soft SUSY
  breaking parameters belongs to the ansatz in Ref.~\cite{KMY2} that
  automatically extends a fine-tuning in the superpotential to the
  full theory.} The scalar masses are ${\cal O}(m_{3/2}^2)$ and depend on
the form of the K\"ahler potential up to ${\cal O}(z^2/M_*^2)$.  For instance,
a term $z^* z \phi^* \phi/M_*^2$ in the K\"ahler potential gives
additional contributions to the $\phi$
scalar mass squared if
$z$ has an $F$-component expectation value.

% So far the discussion has been completely general.
% as long as there are no Planck-scale expectation values
% in the hidden sector.
% {}From this point on, we distinguish models that have gauge singlets
% from those which do not.
If there were a gauge-singlet field with an $F$-component VEV
%$F_z = W_z + {\cal O}(\mu_{SUSY}^3/M_*) = {\cal O}(\mu_{SUSY}^2)$,
$F_z = {\cal O}(\mu_{\rm SUSY}^2)$,
it could be used to generate gaugino masses in the observable sector
of the same order of magnitude as the other soft SUSY breaking
parameters by coupling it to the gauge kinetic function:
\begin{equation}
  \int d^2 \theta \frac{z}{M_*} \tr W^\alpha W_\alpha + {\rm h.c.}
\label{secon}
\end{equation}
This operator cannot appear if the model does not contain singlets,
and the standard conclusion is that the leading contribution to the
gaugino mass in such models comes from higher-dimensional operators,
and is therefore $\mu_{\rm SUSY}^3 / M_*^2 \sim 1$~keV or smaller.
Even if a model does contains singlets, the operator in
Eq.~(\ref{secon}) may be forbidden by symmetries, such as
a $U(1)_R$ symmetry.
% If, however, the hidden sector model does not contain such a singlet,
% the gaugino masses are necessarily extremely suppressed.
% This is also
% the case even in models with gauge singlets if the global symmetry of
% the model forbids the coupling above, such as O'Rafeartaigh model with
% a global $U(1)_R$ symmetry.  If the lowest order invariant is
% quadratic (which can happen if the model is vector-like such as in
% \cite{IY2,IT1}), the gaugino masses are of the order of
% $\mu_{SUSY}^3/M_*^2 \sim {\cal O}(1)~\mbox{keV}$, and the suppression is even
% more severe in chiral models where the lowest order invariant is at
% least cubic.

In fact, this has been regarded as one of the most serious problems in
models of dynamical SUSY breaking in the hidden sector,
since most of these models do not contain gauge
singlets.
One possibility is to use vector-like models of SUSY breaking
with gauge-singlet fields having a non-generic superpotential \cite{IY1}.
%and do not insist on global symmetry to drop
%unwanted terms in the superpotential \cite{IY1}.
%Or one can carefully choose the model of dynamical supersymmetry
%breaking such that the presence of gauge singlet fields do not spoil the
%supersymmetry breaking and they acquire $F$-components at the
%tree-level \cite{Nelson}.
Another possibility is to couple singlets to a model with dynamical
SUSY breaking in such a way that SUSY is not restored and the
singlets aquire $F$ components \cite{Nelson}.
Another proposal is to use a mechanism similar to the messenger $U(1)$
\cite{DNS,DNNS} to generate expectation values for the $F$-component
of a gauge singlet fields at two-loop order \cite{BKN}.
These proposals show that gaugino masses can be generated at order
${\cal O}(M_*^{-1})$ in models with singlets, but it remains true
%Even though all of these mechanisms are
%possible, they are by no means generic.
that a generic model of dynamical SUSY breaking appears to give
extremely small gaugino masses.

%It was suggested in the literature that one may generate gaugino masses
%at the one-loop order $\simeq {\cal O}(g^{2}/16\pi^{2}) m_{3/2}$ by
%integrating
%out massive vector-like chiral superfields with a bilinear
%supersymmetry breaking mass term $B = -m_{3/2}$ \cite{DP,BD}.  Indeed,
%a naive calculation of the one-loop diagram generates a gaugino mass
%$m_{\lambda} = - \frac{g^{2}}{16\pi^{2}} T_F m_{3/2}$ where $T_F$ is
%the total Dynkin index of the (reducible) representation.
%%defined by $\mbox{Tr} T^a T^b = T_F\delta^{ab}$.
%However, this misses the problem pointed out by
%\cite{DM} and repeated in the following.  When one integrates out a
%massive degree of freedom, one should be able to write down the
%(local) low-energy effective field theory.  Then the generated gaugino
%mass (if any) should be written in terms of local operator.  However,
%as we discussed earlier, such an operator gives rise to extremely
%suppressed gaugino masses.  Therefore, we should not obtain a gaugino
%mass at ${\cal O}(g^{2}/16\pi^{2}) m_{3/2}$.

%Another proposal in the literature deserves special mention because it
%has some similarities with the results presented here.
%It was claimed in Refs.~\cite{DP1,DP2,BD}
Another natural possibility would be that a gaugino mass is
generated at 1-loop order from massive vector-like chiral superfields
with a SUSY-breaking mass term $B = -m_{3/2}$ from SUGRA.
Indeed, a direct calculation appears to confirm this, giving a gaugino mass
$\sim g^2 m_{3/2} / (16 \pi^2)$.
However, it was pointed out in Ref.~\cite{DM} that one should be able to
integrate out the massive vector-like matter, and write an effective
low-energy theory in which the gaugino mass (if any) appears as a local
operator.
But we have seen that all such operators give gaugino masses suppressed by
additional powers of $M_*^{-1}$, and so the effect %claimed by
%Refs.~\cite{DP1,BD}
should be absent.
In fact, Ref.~\cite{DM} showed that a careful one-loop calculation using
Pauli--Villars regulator gives a vanishing gaugino mass, because
the Pauli--Villars regulator also has a SUSY-breaking mass from SUGRA
that precisely cancels the contribution from the vector-like multiplet.
The fact that the regulator necessarily breaks SUSY in models of
SUSY breaking in the hidden sector is one of the ways of deriving
the results we present below.

%One may worry about the validity of the argument because a naive
%one-loop calculation indeed generates a gaugino mass.  This
%apparent paradox was resolved using the Pauli--Villars regulator in
%\cite{DM}.  The point is that when one integrates out the massive
%degree of freedom, the theory is not fully regulated.  Even though the
%one-loop diagram that generates the gaugino mass is finite, this
%diagram is related to the chiral anomaly by supersymmetry, and hence
%the issue of the regularization cannot be by-passed.  One can employ
%the Pauli--Villars regulator for this purpose.  When one couples both
%the physical field and the Pauli--Villars
%regulator to supergravity, both of them
%acquire bilinear supersymmetry breaking mass term $B = -m_{3/2}$.
%Integrating out both the regulator and the physical field generates
%gaugino masses, with the opposite sign.  Since the one-loop diagram is
%independent of the (supersymmetric) mass of the particle (as long as
%they are much heavier than $m_{3/2}$), the cancellation is exact.
%Therefore one cannot obtain gaugino masses by integrating out heavy
%degrees of freedom in the observable sector.

\section{Gaugino Mass from Light Multiplets}

% In this section, we show that the gaugino masses are generated in the
% observable sector proportional to the $\beta$-function for the
% corresponding gauge groups.  This mechanism does not depend on the
% details of the hidden sector; in particular, it does not need any
% gauge-singlet fields with expectation values in $F$-component.

In this section, we show that in models with no gauge singlets
the gaugino masses in the observable sector are
proportional to
$\beta m_{3/2}$, where $\beta$ is the beta function for the corresponding
gauge group. In a similar way, the $A$-terms are proportional to the
anomalous dimension of the corresponding Yukawa coupling.%
\footnote{For gaugino masses this result holds \emph{both} for the holomorphic
and `1PI' definitions of the gaugino mass, while for the $A$ terms
there is only a 1PI definition.}
The key point in our analysis is that there is no local operator
that can give a gaugino mass or $A$ term proportional to $1/M_*$.
This implies that the ${\cal O}(M_*^{-1}) = {\cal O}(m_{3/2})$
contributions to these
quantities (if present) are completely finite and calculable in the low-energy
effective theory, since there is no counterterm for the effect.
We will establish a nonzero quantum contribution to the gaugino masses and
$A$ terms using several different methods.
First, we show by explicit calculation that the effect arises when we use
locally supersymmetric regulators for matter loops in the observable
sector.
% Then we show that the result is consistent when crossing
% effective field theory thresholds.
Then we give a general operator analysis that shows that the effect
appears in the 1PI effective action as a direct consequence of local
supersymmetry.
Finally, we show that the effect can be directly understood in terms
of the conformal anomaly multiplet.

\subsection{Explicit Calculations}
% The key point of the analysis is to appropriately regularize the gauge
% theory in the observable sector.  As shown in the previous section,
% one runs into erroneous conclusion if the theory is not appropriately
% regularized.  We used Pauli--Villars regularization in particular to
% show that heavy chiral multiplets do not generate gaugino mass
% following \cite{DM}.  By further extending the analysis to the case
% with light matter and gauge multiplets, we find that gaugino masses
% are generated without gauge-singlet fields.  We employ a particular
% regularization using finite $N=2$ theories and show that $m_{\lambda}
% = \frac{g^{2}}{16\pi^{2}} b_{0} m_{3/2}$.  Then we try to check the
% regulator independence of the result by using a few other
% regularizations.  Next we find a non-trivial consistency check of the
% general result by making some of the fields massive.  In the end we
% discuss the interpretation of the result based on the superconformal
% anomaly multiplet.
We begin by explaining how gaugino masses %and $A$ terms
are generated at the quantum level when we carefully regulate the theory.
(We will discuss $A$ terms only in the next section, where we give
more general arguments.)
Since we are not interested in loops of SUGRA fields, it is
sufficient to regulate matter and gauge loops in the presence of a fixed
SUGRA background.

We would like to write an effective theory for the observable sector with
the hidden sector fields integrated out.
Note that we cannot integrate out the full SUGRA multiplet,
since the graviton is massless.
However, the contribution to the gaugino mass and $A$-terms we are
interested in are ${\cal O}(M_*^{-1})$, while the
exchange of propagating supergravity fields is ${\cal O}(M_*^{-2})$.%
\footnote{
This is true even if we take into account the constant term in the
superpotential proportional to $M_*$ that is needed to cancel the cosmological
constant.
}
At order ${\cal O}(M_*^{-1})$ we can therefore drop the propagating
SUGRA fields and keep only the VEV of the scalar
auxiliary field of the SUGRA multiplet proportional to
$m_{3/2}$.%
%\footnote{
%Since we are assuming that there are no gauge singlets, there are
%no higher-dimension operators connecting the hidden and observable
%sectors at ${\cal O}(M_*^{-1})$.
%}

The ${\cal O}(M_*^{-1})$ terms have a very simple form, which is
easiest to understand using the superconformal calculus formulation of
SUGRA \cite{sugra}.
In this formulation, one first constructs an action invariant under local
superconformal transformations, and then breaks local superconformal symmetry
explicitly down to local super-Poincar\'e symmetry to define the lagrangian.
Every field is assigned a Weyl weight (scaling dimension),
and conformal invariance is broken explicitly by a `compensator' field
${\cal E}$ with Weyl weight $+1$.
${\cal E}$ is taken to have value
${\cal E} = 1 + H \theta^2$, where $H$ is the auxiliary scalar field of
SUGRA, with $\langle H \rangle = m_{3/2}$.
The important feature for our purposes
is that $H$ appears only in ${\cal E}$, and so the
$H$ dependence is determined entirely by dimensional analysis.
This gives the ${\cal O}(m_{3/2})$ SUGRA effects in a very simple form:
\begin{eqnarray}
{\cal L}
\!\!&=&\!\! \int \!\! d^2\theta d^2\bar{\theta}
\sum_\Omega \left[1 + {\textstyle\frac{1}{2}} \left(2 - \dim(\Omega)
\right) m_{3/2}
(\theta^2 + \bar{\theta}^2) \right] \Omega
\nonumber\\
&&
\label{eq:SUGRA}
+\, \biggl( \int \!\! d^2\theta \sum_\Xi
\left[ 1 + (3 - \dim(\Xi)) m_{3/2} \theta^2 \right] \Xi
+ \hbox{\rm h.c.} \biggr) + {\cal O}(m_{3/2}^2),
\end{eqnarray}
where `$\dim$' denotes the total mass dimension of fields and derivatives
in the operators $\Omega$ and $\Xi$
({\it i.e.} the coupling constants do not contribute to the dimension).
The close connection between the coefficient of the linear term in
$m_{3/2}$ and the dimension is a key ingredient in our results.
Note that % $\dim(W^\alpha W_\alpha) = 3$, so
we reproduce the well-known fact that there is no
${\cal O}(m_{3/2})$ contribution to
the gaugino mass or trilinear scalar couplings
in the local lagrangian of supergravity.

% We employ the regularization of gauge theories using finite $N=2$
% theories.\footnote{These theories are known to be finite even
% non-perturbatively, but this is not important to our perturbative
% analysis.}  To regulate a $SU(N_{c})$ theory with $N_{f}$ vector-like
% quarks $Q$ in the fundamental and $\bar{Q}$ in the anti-fundamental
% representations, we further add $2N_{c} - N_{f}$ vector-like quarks
% and an adjoint multiplet $\phi$ together with the superpotential $W =
% \sqrt{2} \bar{Q} \Phi Q$.  Here, the kinetic term of the adjoint field
% is normalized as $1/g^{2}$ because the $N=2$ supersymmetry
% requires the same normalization as the kinetic term of the gauge
% multiplet.  By adding mass terms to the additional quarks and the
% adjoint multiplet, the theory remains ultraviolet finite, while the
% low-energy theory contains only $N_{f}$ vector-like quarks as wished.

The universal nature of the $m_{3/2}$ dependence given above means that
if we regulate the theory in a supersymmetric manner, the regulator
will depend on $m_{3/2}$ in a well-defined way.
This SUSY breaking in the regulator sector will induce finite
SUSY-breaking effects at loop level that give the contribution to the
gaugino mass we are discussing.

For example, we can regulate SUSY QCD with $F \le 2N$ flavors by
imbedding it in a finite ${\cal N} = 2$ theory.%
\footnote{
These theories are known to be finite even non-perturbatively, but this
is not important for our analysis.
}
We can add $2N - F$ vector-like quarks and an
adjoint chiral multiplet $\Phi$ together with the superpotential
$W = \sqrt{2} \bar{Q} \Phi Q$ to obtain a finite ${\cal N} = 2$ theory.
Adding mass terms for $\Phi$ and
the extra quarks breaks ${\cal N} = 2$ SUSY down to ${\cal N} = 1$
maintaining finiteness of the theory, while only the desired
degrees of freedom survive at low energy.
% We now calculate the gaugino mass in the low-energy theory by
% integrating out the massive adjoint and additional vector-like
% quarks.
We then compute the physical gaugino mass in this theory at 1 loop,
including the contribution from the regulator fields.\footnote{See
the appendix for an explanation of why the auxiliary equation of
  motion does not produce an additional contribution to the gaugino
  mass via the Konishi anomaly.}
% Because the $B$-term is universally $-m_{3/2}$, the massive
% adjoint contributes $-(g^{2}/16\pi^{2}) N_{c} m_{3/2}$, while the
% vector-like quarks $- (g^{2}/16\pi^{2}) (2 N_{c} - N_{f})
% m_{3/2}$.
Because the $B$-term for all massive fields is $-m_{3/2}$, the
adjoint contributes $(g^{2}/16\pi^{2}) N m_{3/2}$ at one loop,
while the  additional vector-like quarks contribute
$(g^{2}/16\pi^{2}) (2 N - F) m_{3/2}$.
(These contributions can be viewed as gauge-mediated SUSY
breaking~\cite{DNS,DNNS,grrev} from the regulator sector.)
The result at one loop is therefore
\begin{equation}
        m_{\lambda} = \frac{g^{2}}{16\pi^{2}} (3 N - F) m_{3/2} .
        \label{eq:result}
\end{equation}
% This is the main result of the paper.
% It is important to note that
% the combination $3 N_{c} - N_{f}$ is what appears in the one-loop
% beta-function in the low-energy theory.  Note also that this result
% holds independent of the renormalization scale because $m_{\lambda}$
% and $g^{2}$ run in the same way.  {\bf Do we claim that this result is
%   exact for `Wilsonian' gaugino mass?  I suspect it is true, but
%   this is you guys' expertise!}
note that the result is proportional to the 1-loop beta function
coefficient $b_0=3N-F$  of the low-energy theory.
We will show in the next subsection that this result generalizes to
arbitrary theories (with arbitrary regulators)
and to all orders in perturbation theory.

% We believe that the above result Eq.~(\ref{eq:result}) is valid
% independent of the particular regularization employed.  The reason is
% that, as we discussed in the previous section, any massive degrees of
% freedom cannot contribute to the gaugino mass because of the operator
% analysis in the local low-energy effective theory.  Therefore the
% gaugino mass must be determined by massless particle content only, and
% our result indeed satisfies this criterion.

% To verify that the result is regulator independent, we check a few
% other regularizations.  One is Pauli--Villars regulator that can be
% used for contributions of the vector-like chiral multiplets.  In this
% case, the analysis is precisely the same as for the massive chiral
% multiplets in the previous section \cite{DM}.  By integrating out a
% massive Pauli--Villars regulator field, the contribution to the
% gaugino mass is $- (g^{2}/16\pi^{2}) T_{F} m_{3/2}$.  The negative
% sign reflects the wrong statstics of the Pauli--Villars regulator
% field.  If the physical field is massive, it also produces the same
% contribution with the opposite sign that reflects the correct
% statistics, and they cancel with each other.  However for massless
% chiral multiplet, only the Pauli--Villars field should be integrated
% out and hence the gaugino mass remains uncancelled.  This result
% agrees with the earlier one (\ref{eq:result}) for the contributions of
% the chiral multiplets.

% We can also consider other regulators.
We can also compute the contributions of vector-like chiral multiplets using
Pauli--Villars regularization.
When computing the physical gaugino mass at one loop, the
massive Pauli--Villars fields give a contribution to the gaugino mass of
$-g^2 T_r m_{3/2} / (16\pi^2)$, where $T_r$ is the index of the
representation and the minus sign comes from the `wrong' statistics
of the Pauli-Villars field.
Again this is consistent with Eq.~(\ref{eq:result}).

% The contribution of the gauge multiplet can be obtained using $N=4$
% regularization as well.  In this case, we introduce three chiral
% multiplets in the adjoint representation, and make all of them
% massive.  The low-energy gaugino mass is obtained by integrating out
% all three adjoint multiplets with the universal $B=-m_{3/2}$:
% $m_{\lambda} =  (g^{2}/16\pi^{2}) 3 N_{c} m_{3/2}$, where the factor
% of three comes from three adjoints.  This agrees with the contribution
% of the gauge multiplet in Eq.~(\ref{eq:result}).

The contribution of the gauge multiplet can also be obtained by imbedding
the theory into an ${\cal N} = 4$ theory.
We introduce 3 additional chiral multiplets $\Phi_j$
in the adjoint representation with superpotential
$W = \sqrt{2} \tr(\Phi_1 [\Phi_2, \Phi_3])$,
and add mass terms for the $\Phi$'s to break the theory down to
${\cal N} = 1$.
At one loop, the regulator fields give a
contribution to the gaugino mass $-3 g^2 N m_{3/2} / (16\pi^2)$,
where the factor of 3 comes from the 3 adjoints.

% In principle, one should be able to obtain the same result from other
% more technically involved regularization methods as well, such as
% higher derivative regularization \cite{FS}, or the infinite tower of
% Pauli--Villars regulator \cite{Slavnov}.  These methods can be applied
% to chiral matter content and/or the case with the most general Yukawa
% couplings as well.  We, however, present non-trivial consistency
% checks of the result as an additional evidence of the regulator
% independence rather than going into more complicated regularization
% methods.  In these consistency checks, it is crucial that the result
% (\ref{eq:result}) is proportional to the beta-function.

Finally we can consider
dimensional reduction \cite{siegel}, in which the $d$-dimensional
superconformal invariance modifies Eq.~(\ref{eq:SUGRA}) to
\begin{eqnarray}
{\cal L}
\!\!&=&\!\! \int \!\! d^2\theta d^2\bar{\theta}
\sum_\Omega \left[1 + {\textstyle\frac{1}{2}} \left(d - 2 - \dim(\Omega)
\right)
m_{3/2}
(\theta^2 + \bar{\theta}^2) \right] \Omega
\nonumber\\
&&
\label{eq:dSUGRA}
+\, \biggl( \int \!\! d^2\theta \sum_\Xi
\left[ 1 + (d - 1 - \dim(\Xi)) m_{3/2} \theta^2 \right] \Xi
+ \hbox{\rm h.c.} \biggr) + {\cal O}(m_{3/2}^2),
\end{eqnarray}
where we define the Weyl weights of fundamental superfields to be equal
to their mass dimension in $d = 4 - \epsilon$ dimensions.
Note that the vector superfield is dimensionless, and so the gauge
biliner $W_\alpha W^\alpha$ has dimension $3$ for all $d$.
Therefore  by Eq.~(\ref{eq:dSUGRA}) the bare gauge kinetic term is
\begin{equation}
{\cal L}_{\rm gauge}
= \int\!\!d^2\theta\, (1-\epsilon m_{3/2} \theta^2) {1 \over 4 g_0^2}
W^\alpha_A W_{\alpha A},
\label{gaugekin}
\end{equation}
where $g_0^2$ is the bare coupling and $A$ is a gauge index.
This lagrangian contains a bare gaugino mass equal to
$-m_{3/2}\epsilon$ that combines with the $1/\epsilon$ terms in the bare
gauge coupling to give a finite gaugino mass.
At one loop, we obtain
\begin{equation}
m_{\lambda} = \left(  m_{3/2} \epsilon \right)
\left( \frac{b_0 g^2}{16\pi^2} \frac{1}{\epsilon} \right) .
\end{equation}
% where the second term is the 1-loop gaugino mass counterterm.
% Here, $b$ is the coefficient of the 1-loop gauge coupling RG, so this
% again reproduces the result.
We could consider other supersymmetric regulators, such as
higher-derivative regulators \cite{FS}
or an infinite tower of Pauli--Villars regulators \cite{Slavnov},
but we will stop here.

We can gain additional insight into this result if we note that
the proportionality between the gaugino mass and the beta function
of the low-energy effective field theory is preserved across
effective field theory thresholds.
This can be seen by direct 1-loop calculation, but it follows
more directly from the method of `analytic
continuation into superspace' \cite{GR,AGLR}.
At 1-loop order, the gauge coupling and gaugino mass can be
grouped into a chiral superfield
\begin{equation}
S(\mu) = \frac{1}{2 g^2(\mu)} - \frac{i\Theta}{16\pi^2}
- \frac{m_\lambda(\mu)}{g^2(\mu)} \theta^2,
\label{eq:S}
\end{equation}
and the effects of a threshold at the scale $M$ is calculated using the
1-loop RG equation $\mu d S / d\mu = b / (16\pi^2)$:
\begin{equation}
\label{Seq}
S_{\rm eff}(\mu) = S(\mu_0) + \frac{b}{16\pi^2} \ln\frac{M}{\mu_0}
+ \frac{b_{\rm eff}}{16\pi^2} \ln\frac{\mu}{M}.
\end{equation}
Here $\mu_0 > M$ is the renormalization scale used to define the
fundamental theory, and $\mu < M$ is the renormalization scale in the
effective theory.
In all cases of interest, the scale $M$ can be written as a chiral
superfield.
For example, if we are integrating out a massive
vector-like chiral field, its mass $M$ appears in the superpotential and can
be analytically continued to a full chiral superfield.
The other possibility is that the mass threshold is due to the
VEV of a chiral superfield, which can partially break the gauge
symmetry and/or give mass to some vector-like multiplets.
In all cases, it is easily checked that Eq.~(\ref{Seq}) is correct
in the limit of unbroken supersymmetry.

If $m_{3/2} \ll M$, the theshold at the scale $M$ is approximately
supersymmetric.
In this case,
Ref.~\cite{GR} showed that Eq.~(\ref{Seq}) remains correct
in the presence of SUSY breaking if the $\theta$-dependent components of
$M$ are included.
(There are additional subtleties beyond 1 loop; see Ref.~\cite{AGLR}.)
By Eq.~(\ref{eq:SUGRA}), this amounts to the substitution
$M \rightarrow M ( 1 + m_{3/2} \theta^2)$,
which gives
\begin{equation}
\frac{m_{\lambda,{\rm eff}}(\mu)}{g_{\rm eff}^2(\mu)}
= \frac{m_{\lambda}(\mu_0)}{g^2(\mu_0)}+
 \frac{b_{\rm eff} - b}{16\pi^2} m_{3/2}
= \frac{b_{\rm eff}}{16\pi^2} m_{3/2}.
\end{equation}
Note that this result includes the correct 1-loop RG evolution down to
the scale $\mu$.

To make this more explicit, consider for example $SU(N)$
gauge theory with one flavor  broken down to
$SU(N-1)$ by the Higgs mechanism.
%By integrating out the massive vector-multiplet, we can
% calculate the threshold correction to the gaugino mass.  Starting
% from the gaugino mass in the high-energy $SU(N)$ theory using our
% result in Eq.~(\ref{eq:result}), we obtain the gaugino mass in the
% low-energy $SU(N-1)$ theory, which agrees with the general result
% (\ref{eq:result}) again.  We consider this as a non-trivial
% consistency check and hence gives an additional evidence for the
% result.
%
We take a superpotential $W = \lambda X
 (Q \bar{Q} - v^2)$, where $X$ is a singlet and $Q$, $\bar Q$ are
one flavor in the fundamental of $SU(N)$.
  In the SUSY limit, we find $\langle Q\rangle =
 \langle \bar{Q} \rangle = v$.  In the presence of soft SUSY breaking
 terms, the potential is
 \begin{equation}
  V = \lambda^2 |Q\bar{Q} - v^2|^2 + \lambda^2 (|Q|^2 + |\bar{Q}|^2)
  |X|^2 + 2 m_{3/2} (\lambda X v^2 + \hbox{\rm h.c.}).
 \end{equation}
 We find $\langle X \rangle = - m_{3/2}/\lambda$, and hence $F_Q =
F_{\bar{Q}} = -\lambda \langle X \rangle v =  m_{3/2} v$. The low-energy
effective superfield coupling is
\begin{equation}
S_{\rm eff}(\mu)= {S(\mu)}+ {1\over 16 \pi^2}\ln ({Q\bar Q
\over \mu^2})
\end{equation}
where $S(\mu)$ is the coupling of the high energy theory.
% When the $SU(N)$ breaks to $SU(N-1)$, the matching of the gauge
% couplings are given by
% \begin{equation}
%   \frac{8\pi^2}{g^2_{N-1}} = \frac{8\pi^2}{g^2_N} + \log
%   \frac{Q\bar{Q}}{M^2},
% \end{equation}
% where $M$ is the UV cutoff.
The $F$ components in $Q$ and $\bar{Q}$ modify the gaugino mass by
 \begin{equation}
   \Delta\left( \frac{m_\lambda}{g^2}\right)
 = -\frac{1}{16\pi^2} \left(\frac{F_Q}{Q}
   + \frac{F_{\bar{Q}}}{\bar{Q}} \right) = - \frac{2}{16\pi^2} m_{3/2}.
 \end{equation}
This factor of 2 is the difference in beta-function coefficients,
so the resulting low-energy gaugino mass is precisely what one obtains
with our formula (\ref{eq:result}) applied to the effective $SU(N - 1)$
gauge theory.

\subsection{General Argument}
We have seen that at one loop the gaugino mass is proportional to the
beta function of the low-energy theory.
This strongly suggests that there is a close
connection between the effect we are discussing and the conformal
properties of the theory.
We now give a general argument that shows this connection explicitly,
and generalizes the results of the previous subsection to arbitrary
theories and to all orders in perturbation theory.

The starting point is a definition of the 1PI gaugino mass using an
operator analysis in superspace,
following Ref.~\cite{AGLR}.
A useful definition of the 1PI gauge coupling and gaugino mass can be
obtained by considering the 1PI gauge 2-point function expanded at
short distances (compared to $m_\lambda^{-1}$).
The leading term in the expansion in $1/\Delamb$ is
\begin{equation}
\label{eq:opidef}
\Gamma_{\rm 1PI} = \int \!\! d^4 x \int \!\! d^2\theta d^2\bar{\theta}\,
W^\alpha_A R(\Delamb) \left( -\frac{D^2}{8\Delamb} \right) W_{\alpha A}
+ \hbox{\rm h.c.} + \cdots
\end{equation}
The function $R(\Delamb)$ has a logarithmic dependence on $\Delamb$
that is the source of the 1PI renormalization group.
The identity
\begin{equation}
\int \!\! d^2\theta d^2\bar{\theta}\, W^\alpha_A
\left( -\frac{D^2}{8\Delamb} \right) W_{\alpha A}
= \frac{1}{2} \int \!\! d^2\theta\,
W^\alpha_A W_{\alpha A}
\end{equation}
shows that the leading term in Eq.~(\ref{eq:opidef}) is local in coordinate
space even though it is nonlocal in superspace.
A general operator analysis \cite{AGLR}
can be used to show that all other operators
that contribute to the gauge 2-point function are suppressed by powers
of $1/\Delamb$.
This shows that the superfield $R$ contains the 1PI gauge coupling and
gaugino mass as its lowest components:
\begin{equation}
R(\Delamb = -\mu^2) = \frac{1}{g^2(\mu)} - \left(
\frac{m_\lambda(\mu)}{g^2(\mu)} \theta^2 + \hbox{\rm h.c.} \right)
+ \cdots
\end{equation}
For a more complete discussion (including the meaning of the
$\theta^2 \bar{\theta}^2$ components of $R$) see Ref.~\cite{AGLR}.

We can now write the covariant generalization of Eq.~(\ref{eq:opidef}) in
a SUGRA background using the results quoted in Eq.~(\ref{eq:SUGRA}).
Since $W_\alpha (D^2/\Delamb) W^\alpha$ has dimension 2,
the ${\cal O}(m_{3/2})$ terms are obtained simply by making the replacement
\begin{equation}
R(\Delamb) \rightarrow R(\Delamb
[1 - (m_{3/2} \theta^2 + \hbox{\rm h.c.})]).
\end{equation}
Expanding the terms linear in $m_{3/2}$, we obtain the gaugino mass
\begin{equation}
m_\lambda = \frac{g^2 m_{3/2}}{2} \mu \frac{d R}{d\mu}
= -\frac{m_{3/2}}{2 g^2} \mu \frac{d g^2}{d\mu}=-{\beta(g^2)\over 2
g^2}m_{3/2}.
\label{allorder}
\end{equation}
Note that $g$ and $m_\lambda$ are 1PI renormalized couplings,
defined in a `superfield' scheme where they are the components of a
real superfield.
This result generalizes our previous result to all orders in perturbation
theory.

% The relation between the scaling dimension of operators and
% the ${\cal O}(m_{3/2})$ terms given in Eq.~(\ref{eq:SUGRA}) shows that
% the vanishing of the gaugino mass at the classical level is directly
% related to the fact that it arises from a dimension-4 operator in the
% lagrangian.
% The quantum violation of scale invariance then generates a non-trivial
% gaugino mass.
%
% It is now clear why the naive operator analysis, which gives
% zero gaugino mass, fails at the quantum level. That analysis corresponds to
% rendering SUGRA covariant a {\it local} tree level lagrangian.
% At the quantum level, one deals instead with a  {\it non-local} action,
% which must be embedded in a non-trivial SUGRA background. Now, the funny
% thing
% is that the SUGRA covariant version of non-local operators involves
% also local pieces. Basically this is because the  D'Alembertian of
% rigid supersymmetry
% $\Delamb$  becomes $ \Delamb_{SUGRA} =\Delamb(1-H \theta^2 -H^*\bar
% \theta^2)$
%  so that the SUGRA covariant version of $(\ln \Delamb)$ involves a local
% supersymmetry breaking piece.

This argument shows very directly the connection between the quantum
contribution to the gaugino mass and the conformal anomaly.
The point is that SUGRA covariance relates the ${\cal O}(m_{3/2})$
soft breaking terms to the scaling dimension of the operators in
the SUSY limit.
At tree level, this relation is given in Eq.~(\ref{eq:SUGRA});
our analysis above shows that this relationship is preserved at
the loop level as well, so that the ${\cal O}(m_{3/2})$ terms
depend on the quantum scaling dimension of the operators.
This arises because the physical gaugino mass must be read
off from the 1PI effective action in the presence of a SUGRA background.
SUGRA covariance mandates the replacement
$\Delamb \rightarrow \Delamb[1 - (m_{3/2} \theta^2 \hbox{\rm h.c.})]$,
which means that the SUGRA covariant version of $\ln\Delamb$ contains
a local SUSY-breaking piece.

% In a completely analogous manner we can calculate the quantum corrections to
% $A$-type soft terms. For any superpotential interaction involving $n$-fields
% $W=\lambda \phi_1\dots\phi_n$, Eq.~\ref{eq:SUGRA} associates at tree level
%  soft breaking term $V_{\rm soft}=(n-3)m_{3/2}\lambda \tilde
% \phi_1\dots\tilde
%  \phi_n$. The quantum corrections are simply calculated by considering
% the matter kinetic terms in the 1PI action
% \begin{equation}
% \int d^2\theta^2\bar \theta^2 \phi_r^\dagger Z_r\left(
% \Delamb[1-m_{3/2}\theta^2
% -m_{3/2}\bar \theta^2]\right )\phi_r
% \label{matterkin}
% \end{equation}
% at an external momentum $\Delamb= -\mu^2$. By expanding $Z_r$ in $m_{3/2}$
% we find that the $A$-type terms renormalized at a scale $\mu^2$ are just
% \begin{equation}
% A_n(\mu)=\left (n-3-{1\over 2}\sum_1^n \gamma_r(\mu)\right ) m_{3/2}
% \label{aquantum}
% \end{equation}
% where we used the definition of the anomalous dimensions $\gamma_r(\mu)=
% d\ln Z_r(\mu^2)/d\ln \mu$. Notice that the expression in bracket can
% suggestively be written as $n_{quantum}-3$, {\it i.e.} in terms of the
% quantum
% dimension of the chiral operator.  Again, trilinear soft
% terms are just proportional to the Yukawa coupling RG equation
% $A_3=-m_{3/2}d\ln\lambda(\mu)/d\ln \mu$.

We now briefly consider $A$ terms.
For a dimension $n$ term in the superpotential of the form
$W = \lambda \Phi_1 \cdots \Phi_n$, Eq.~(\ref{eq:SUGRA}) gives a
tree-level soft-breaking term
$V_{\rm soft} = (n-3) m_{3/2} \lambda \phi_1 \cdots \phi_n$.
We now read off the quantum corrections to this from the kinetic terms
in the 1PI effective action.
The leading term in the expansion in $1/\Delamb$ is
\begin{equation}
\int\!\! d^2\theta d\bar\theta^2\,
\Phi_r^\dagger Z_r(\Delamb[1- (m_{3/2}\theta^2 + \hbox{\rm h.c.})])
\Phi_r + {\cal O}(m_{3/2}^2).
\label{matterkin}
\end{equation}
The 1PI renormalized wavefunction and $A$ terms can be defined by
appropriate components of $Z_r(\Delamb = -\mu^2)$.
We then find that the $A$-type terms renormalized at a scale $\mu^2$ are
\begin{equation}
A_n(\mu)=\left (n-3-{1\over 2}\sum_{r = 1}^n \gamma_r(\mu)\right ) m_{3/2}
\label{aquantum}
\end{equation}
where
\begin{equation}
\gamma_r(\mu) = \mu \frac{d \ln Z_r}{d\mu}
\end{equation}
is the anomalous dimension.
Notice that the right-hand-side of Eq.~(\ref{aquantum}) is proportional
to the quantum dimension of the chiral operator minus 3.
We see that trilinear soft terms are proportional to the beta function of the
corresponding Yukawa coupling:
\begin{equation}
A_3 = -m_{3/2} \mu \frac{d \ln\lambda}{d\mu}.
\end{equation}

The results we have quoted above are valid to all orders in perturbation
theory, and we make some comments on scheme dependence.
The preceding derivation makes clear that the results hold in any
scheme in which the SUSY-breaking couplings are treated as higher
components of superfield couplings.
In Ref.~\cite{AGLR,jonesexact} it was shown that such a definition is always
possible to all orders in perturbation theory, and this class of schemes
were called `superfield coupling schemes'.
In the literature there are many examples of `exact' results for soft terms
whose derivation is based on the all-orders beta function of Novikov,
Shifman, Vainshtein and Zakharov (NSVZ) \cite{NSVZ}.
If these results truly depended on the precise form of the all-orders
beta function, they would be valid only in the NSVZ scheme where the
beta function takes the form of Ref.~\cite{NSVZ}.
However, the study in Ref.~\cite{AGLR} shows that these results are in
fact valid in any superfield coupling scheme.
One example of such an `exact' relation is \cite{hisano}
\begin{equation}
{g^2 m_\lambda\over \beta}
+ {1\over b_0} \sum_r T_r \left(
\ln Z_r(\mu^2)|_{\theta^2} - {\gamma_r g^2 m_\lambda
\over \beta} \right) = {\rm RG\ invariant}.
\label{rgi}
\end{equation}
In the class of theories we are considering, the second term on the
left-hand-side vanishes by the results for $A$ terms derived above.
We then obtain
\begin{equation}
{g^2m_\lambda \over \beta} = -{\textstyle {1 \over 2}}
m_{3/2} = {\rm RG\ invariant}.
\end{equation}
It was pointed out in Ref.~\cite{hisano} that
this relation is in general valid
only in the absence of Yukawa interactions.
Our results imply that this relation is true in minimal supergravity
even in the presence of other interactions, and hold in any superfield
scheme.

\subsection{Superconformal Anomaly Multiplet}

In this subsection, we present an alternative argument which justifies
Eqs.~(\ref{allorder}) and (\ref{aquantum}).
The argument assumes the existence of a
manifestly supersymmetric and holomorphic regularization, as those
based on finite ${\cal N}=2$ or ${\cal N}=4$
theories.  However, it does not depend
on the details of the regularization procedure.

In an explicitly regulated theory, the ultraviolet cutoff is
provided either by the mass of the regulators ({\it
e.g.}\/, Pauli--Villars fields or extra adjoint and quark fields in
${\cal N}=2$ regularization) or by the inverse mass scale
of the higher-dimensional terms ({\it e.g.}\/,
higher-derivative regularization), or both.  For our purposes, we refer to
the ultraviolet cutoff generically by $M$. The assumption of
a holomorphic regularization implies that
the cutoff $M$ can be regarded as a chiral
superfield spurion appearing in the superpotential.
{}From Eq.~(\ref{eq:SUGRA}) it is easy to see that the
effect
of supersymmetry breaking is a simple replacement $M \rightarrow
M(1+m_{3/2}\theta^{2})$, independent of details of the regularization
procedure.  Because of manifest holomorphy, the dependence on the
cutoff $M$ fixes the effect of supersymmetry breaking at ${\cal O}(m_{3/2})$.

The Wilsonian renormalization group invariance states that one can
change the cutoff $M$ without changing low-energy physics as long as
one changes the bare parameters in the Lagrangian in a specific
manner.  To be explicit, the statement is that for any (physical)
correlation function $G$
\begin{equation}
        M\frac{d}{dM} G = \left(M \frac{\partial}{\partial M}
        + \frac{b_{0}}{16\pi^{2}}
        \frac{\partial}{\partial S} + \sum_{i} M \frac{d \ln Z_{i}}{dM}
        \frac{\partial}{\partial \ln Z_{i}}\right) G = 0
        \label{eq:RGEinv}
\end{equation}
where $S$ is defined in
Eq.~(\ref{eq:S}).  Here, the index $i$ runs
over all chiral superfields in the theory, and $b_{0}$ is the one-loop
beta function coefficient of the gauge coupling constant.  We also
assumed that there is no dimensionful coupling constant in the theory;
if any, one can trivially extend the analysis by including the
dimensionful terms as an explicit breaking of scale invariance to be
added to
the right-hand side of Eq.~(\ref{eq:RGEinv}).
Note that $M d \ln Z_{i} / dM = \gamma_{i}$
to the lowest order in $\theta$, $\bar{\theta}$.

To find the effect of the replacement $M \rightarrow
M(1+m_{3/2}\theta^{2})$ in the presence of supersymmetry breaking in
the hidden sector, one can use the renormalization group invariance and
integrate Eq.~(\ref{eq:RGEinv}) from a constant $M$ to
$M(1+m_{3/2}\theta^{2})$.  This is equivalent to the technique of `analytic
continuation to superspace' \cite{GR,AGLR}.
The derivative with respect to the cutoff
$M$ inserts the trace of the energy-momentum tensor
$\Theta_{\mu}^{\mu}$ to the correlation function.  It has been known for
more than two decades \cite{Bruno} that the trace of
the energy-momentum tensor belongs to a chiral superfield $\Phi$ called
`anomaly-multiplet' whose $F$-component is $\Theta_{\mu}^{\mu} + i
\frac{3}{2} \partial_{\mu} j_{R}^{\mu}$, where $j_{R}^{\mu}$ is the
$U(1)_{R}$ current.  From the above $M$ derivative in the Wilsonian
effective action, we get
\begin{equation}
        \Phi = \frac{b_{0}}{8\pi^{2}} W_{\alpha} W^{\alpha} + \sum_{i}
        \frac{\gamma_{i}}{16\pi^{2}} \bar{D}^{2} (\phi_{i}^{\dagger} e^{V}
        \phi_{i}).
\end{equation}
Note that the first term can be fixed by the $U(1)_{R}$ anomaly, while
the second term gives a total derivative to the imaginary part of
$\Phi$ and hence cannot be determined from the $U(1)_{R}$ anomaly.
One can easily derive this from Eq.~(\ref{eq:RGEinv}) by noting that
derivatives with respect to $S$ and $\ln Z_{i}$ pulls down the
$W_{\alpha} W^{\alpha}$ and
$\bar{D}^{2}(\phi_{i}^{\dagger}e^{V}\phi_{i})$ operators from the
action in the path integral.  This equation is exact to all orders,
once the one-loop
gauge beta function $b_{0}$ is used.

Now it is easy to see that the integration of Eq.~(\ref{eq:RGEinv})
from a constant $M$ to $M(1+m_{3/2}\theta^{2})$ produces the gaugino
mass and the $A$-terms. The lowest component of the anomaly multiplet
$\Phi$ is the sum of the gaugino-bilinear $\lambda_{\alpha}
\lambda^{\alpha}$, and the operator $F_{i}^{*} A_{i}$ which gives the
$A$-terms upon solving the auxiliary equations of motion for
$F_{i}$.  Here, $A_{i}$
($F_{i}$) is the lowest (highest) component in the chiral superfield
$\phi_{i}$.  This  immediately justifies Eqs.~(\ref{allorder})
and (\ref{aquantum}).

The above argument leads to Eq.~(\ref{allorder}) at the one-loop
level, which is exact in the `holormorphic' definition of the gauge
coupling constant and the gaugino mass employed here but not in the
`canonical' definition which admits a more direct physical
interpretation.  The justification of Eq.~(\ref{allorder}) requires an
additional step to go from the `holomorphic' definition to the
`canonical' definition by changing the normalization of the chiral
and gauge multiplets to the canonical normalization.  This rescaling
of the vector multiplet induces an anomalous Jacobian in the Fujikawa
measure which changes both the gaugino mass and the gauge beta
function in the same manner \cite{hisano}.\footnote{This is not the
way it was discussed in \cite{hisano}.  The explanation here is based
on an extension of the analysis in \cite{AHM}.  In a manifestly
supersymmetric calculation, the contribution of the rescaling of the
gauge multiplet comes from the Konishi anomaly of the $b$-ghost chiral
superfield \cite{AHM}, by going from the original Lagrangian $\int
d^{4} \theta (S+S^{\dagger})\bar{b}b$ to the canonical one.}

\section{Phenomenology}
% The main difference of our framework from the conventional hidden
% sector models is in the soft supersymmetry breaking parameters.  Even
% though the detailed phenomenology is beyond the scope of this paper,
% in this section we briefly discuss the important differences.
In absence of singlet fields
in the hidden sector, we have seen that
gaugino masses are generated, but they turn out to be
of order $\alpha m_{3/2}$ rather than $m_{3/2}$.
Since we expect squark and slepton masses to be of order $m_{3/2}$, the
phenomenology is quite different from conventional hidden sector models.

Before entering into this discussion it is worth questioning whether
the scalar masses are necessarily of order $m_{3/2}$ without additional
suppressions.
Unlike the gaugino masses and $A$ terms, the
scalar masses and $B\mu$ terms are not calculable in the low-energy
effective theory due to the presence of the counterterms
\begin{equation}
\int\!\! d^2\theta d^2\bar\theta\,
\frac{z^\dagger z}{M_*^2} Q^\dagger Q
\quad
\int\!\! d^2\theta d^2\bar\theta\,
\frac{z^\dagger z}{M_*^2} H_u H_d,
\end{equation}
where $z$ are hidden-sector fields, $Q$ are observable-sector
matter fields, and $H_{u,d}$ are Higgs fields.
The coefficients of these terms can be adjusted so that the scalar masses
and $B\mu$ terms are of order $\alpha^2 m_{3/2}^2$ rather than $m_{3/2}^2$.
This appears to be a fine-tuning of order $\alpha^2 \sim 10^{-4}$, but it
is possible that it could be the consequence of
a more fundamental theory such as string theory.
(For example, the scalar masses are naturally suppressed in
`no-scale' models \cite{LN,Heisenberg}.)
More generally,
it is worth noting that if the counterterms are
chosen to make all soft terms of the same order, there is no fine-tuning
evident in the low-energy effective theory below the Planck scale.
In such a theory, the main differences with conventional hidden-sector
models are that the gravitino is much heavier than the other superpartners,
and that gaugino masses satisfy the specific relations discussed below.

We now turn to the phenomenological consequences of the (probably
more likely) scenario in which
scalar masses are of order $m_{3/2}$.
% We have shown that gaugino masses are generated by
% the light fields in the observable sector. However, with respect to
% the gravitino mass $m_{3/2}$, they are
% suppressed by the one-loop factor because they are anomaly effects.
In the case of the minimal supersymmetric extension of the Standard Model,
the gaugino masses at the weak scale are
\begin{eqnarray}
M_1 &=& \frac{11\alpha}{4\pi\cos^2\theta_W} m_{3/2}=8.9\times 10^{-3}m_{3/2},
\label{m1c} \\
M_2 &=& \frac{\alpha}{4\pi\sin^2\theta_W} m_{3/2}=2.7\times 10^{-3}m_{3/2},
\label{m2c} \\
M_3 &=& -\frac{3\alpha_s}{4\pi} m_{3/2}=-2.6\times 10^{-2}m_{3/2}.
\end{eqnarray}
Electroweak gaugino masses receive also contributions from finite
one-loop diagrams
with Higgs and Higgsino exchange. If the supersymmetric Higgs mass $\mu$ is
of the same order of $m_{3/2}$, this contribution is comparable to those
of Eqs.~(\ref{m1c})--(\ref{m2c}). In the limit in which $M_W$ is much smaller
than both $\mu$ and the pseudoscalar Higgs mass $m_A$,
the total result for
the electroweak gaugino masses becomes
\begin{eqnarray}
M_1 &=& \frac{\alpha}{4\pi\cos^2\theta_W}
m_{3/2}~\left[11-f(\mu^2/m_A^2)\right]
,\\
M_2 &=& \frac{\alpha}{4\pi\sin^2\theta_W} m_{3/2}\left[1-f(\mu^2/m_A^2)\right]
,
\end{eqnarray}
\begin{equation}
f(x)=\frac{2x\ln x}{x-1}.
\end{equation}
The present LEP bound on the chargino mass requires $M_2\gtrsim M_W$. This
translates into a lower bound on the gravitino mass $m_{3/2}$ of about
30 TeV for $\mu^2/m_A^2=1$. This bound decreases for larger values
of $\mu^2/m_A^2$ and it is about 8 TeV for $\mu^2/m_A^2=8$. The gluino mass is
heavier than 200 GeV as long as $m_{3/2}>8$ TeV.
As mentioned above,
the scalar masses are expected to be of the same order of magnitude
as the gravitino mass.
This then implies somewhat
dishearteningly large squark and  slepton masses, and requires considerable
fine-tuning in the electroweak symmetry breaking.

For $\mu^2/m_A^2\gtrsim 3$, we find
$|M_1|<|M_2|$ and an almost pure $B$-ino is likely
to be the lightest supersymmetric particle (LSP). In this case, the LSP
relic abundance overcloses the Universe. For instance, assuming that
the three
families of sleptons are degenerate with mass $m_{\tilde \ell}$ and the
squarks are heavier, the LSP contribution
to the present energy density, in units of the critical density, is
\begin{equation}
\Omega_{\rm LSP} h^2 \simeq 90 \left( \frac{\rm 100~GeV}{m_{\chi^0}}\right)^2
\left( \frac{m_{\tilde \ell}}{\rm TeV}\right)^4,
\end{equation}
where $h$ is the Hubble constant in units of 100 km s$^{-1}$ Mpc$^{-1}$.
The problem of the large $\Omega_{\rm LSP}$ could be resolved if some sfermion
masses or $\mu$ are much smaller than the typical scale $m_{3/2}$. The LSP
annihilation cross section can then be increased either by the light
sfermion exchange in the $t$-channel or by the $s$-channel $Z$ exchange
induced by gaugino-higgsino mixing. A more radical solution is to invoke
early LSP decay caused by some $R$-parity violation in the theory.

For $\mu^2/m_A^2\lesssim 3$, we find the more unconventional possibility
that the
$W$-ino is lighter than the $B$-ino.
The mass splitting between the neutral and charged $W$-inos
belonging to the same $SU(2)$ triplet is induced by electroweak breaking,
but occurs (both at the classical and the quantum level) only at order $M_W^4$.
In the limit $\mu \gg M_{1,2}, M_W$, the charged and neutral $W$-ino masses
are
\begin{equation}
m_{\chi^\pm} = M_2-\frac{M_W^2}{\mu}\sin 2\beta +
\frac{M_W^4}{\mu^3}\sin 2\beta ,
\label{chipm}
\end{equation}
\begin{equation}
m_{\chi^0}= M_2-\frac{M_W^2}{\mu}\sin 2\beta
-\frac{M_W^4\tan^2\theta_W}{(M_1-M_2)\mu^2}\sin^2 2\beta ,
\label{chi0}
\end{equation}
where $\tan\beta$ is the ratio of Higgs vacuum expectation values.
{}From Eqs.~(\ref{chipm})--(\ref{chi0}) we infer that
the neutral $W$-ino state is the LSP. $W$-ino annihilation
in the early Universe is very efficient, since two neutral $W$-inos can
produce $W$ boson pairs and charged and neutral $W$-inos can produce
fermion pairs via $W$ exchange. Neglecting for simplicity the co-annihilation
channels, we find
\begin{equation}
\Omega_{\rm LSP} h^2 \simeq 5\times 10^{-4} \left( \frac{m_{\chi^0}}{\rm
100~GeV}\right)^2.
\end{equation}
The neutralino does not cause any problem with relic overabundance, but
cannot be used as a cold dark matter candidate \cite{MNY}.

The chargino search at LEP is more difficult
in the case of a pure $W$-ino LSP than in the case of $B$-ino LSP
\cite{Gunion}.
Because of the small mass difference between $\chi^\pm$ and $\chi^0$
(see Eqs.~(\ref{chipm})--(\ref{chi0})) chargino production leads to
extremely soft final states, and detection could require a photon-tagging
technique (see {\it e.g.} the analysis in Ref.~\cite{drees}). For very
small mass difference, the chargino is so long-lived that it could
be observed through anomalous ionization tracks with little associated
energy deposition in calorimeters. Indeed, the average distance travelled
by a chargino with energy $E$ is
\begin{equation}
L=\left( \frac{\rm GeV}{m_{\chi^\pm} -m_{\chi^0}}\right)^5
\left( \frac{E^2}{
m_{\chi^\pm}^2}-1 \right)^{1/2}\times 10^{-2}\ {\rm cm}.
\end{equation}
This distance could well be macroscopic and exceed the detector size
 when $\mu$ is of the order of
the gravitino mass, since $m_{\chi^\pm} -m_{\chi^0}\sim M_W^3/\mu^2$,
see Eqs.~(\ref{chipm})--(\ref{chi0}). Quasi-stable electromagnetically
charged particles can also be searched at hadron colliders.
Moreover, at hadron colliders, the search can also proceed through
the conventional missing energy signature,
which can now be renforced by an effectively invisible chargino decay,
whenever the chargino decays promptly.
In particular, the most promising missing energy signal comes from
gluinos, which are strongly produced and decay into a pair of quarks and
a neutral or charged invisible $W$-ino.

The most unpleasant feature of the
% Renormalizable Hidden Sector scenario presented in this paper
scenario presented here
is the large hierarchy between scalar and gaugino masses.
Heavy scalars, however, help weakening the problem with
flavor-changing neutral current processes from supersymmetric loop effects.
A certain degree of degeneracy among scalars is still required, but this can
well be a consequence of a flavor symmetry.  This problem is common to
all hidden sector models and not special to this particular
framework. Here it has been alleviated at the price of more fine tuning
in the electroweak breaking condition.

It should be noted that even when supersymmetry breaking is mainly
in the $D$-terms rather than $F$-terms in the hidden sector, the
squark degeneracy is not guaranteed, contrary to the claim of
Ref.~\cite{DP2}.  If there is a large $D$-term expectation value
$D$, the auxiliary equation of motion insures that there is at least one
scalar field which generates $D = z^* Q z \sim {\cal O}(\mu_{SUSY}^2)$, where
$Q$ is the charge of the $z$ field under the gauge generator.  Then
the operator $\int d^4 \theta (z^* e^V z) \phi^* \phi/M_*^2$ gives
contributions to the observable field $\phi$
scalar mass of order $m_{3/2}$, which do not
preserve squark and slepton degeneracy even in models without $F$-term.

An interesting feature of the scenario
is that one can naturally justify the absence of new phases
in the soft breaking terms and therefore satisfy the experimental
constraints on
CP violating processes.
%that the trilinear soft
%supersymmetry breaking parameters vanish identically at the
%tree-level, $A=0$.  This makes the phenomenological constraint from
%the neutron and electron electric dipole moment (EDM) much less
%severe than conventional Polonyi-type hidden sector models.  To see if
%this is indeed the case, one should also study possible CP-violating
%phases in the Higgs sector.
In the minimal supersymmetric model, there are five possible
sources of CP-violating phases: $\mu$, $B\mu$, $M_{i}$
($i=1,2,3$), $A$. The physically observable phases  are only those combinations
that are
invariant under $U(1)_{R}$ and Peccei--Quinn phase rotations. In our framework
there is just one parameter, $m_{3/2}$, that breaks $U(1)_R$
and just one, $\mu$, that breaks PQ, therefore there is no physical phase.
% one can
%eliminate two of the phases.  In our framework, trilinear soft
%supersymmetry-breaking parameters vanish identically at the
%tree-level, $A=0$.
%This leaves only one
%phase: the relative phase between $M_{i}$ and $B$ in the basis where
%$\mu$ is real.  Interestingly enough, $M_{i}$ and $B$
%are both given by $m_{3/2}$, and hence there is no relative phase.
This makes the constraints on neutron and electron electric dipole moments
automatically satisfied.

If the origin of $B\mu$ is different than the universal $B$-term in
Eq.~(\ref{eq:universal}), this feature may be
spoiled.  The question then becomes somewhat model dependent, and
it is connected with the $\mu$ problem that will be addressed in the
next section.
%As long as the sector responsible for the $\mu$-term
%does not contain a CP-violating phase, the absence of the CP-violating
%effects remains.

Our framework does not address the structure of the scalar masses, which
depends on the specific form of the K\"ahler potential,  and
therefore nothing can be said about possible imaginary parts of the squark
mass matrix. These phases can lead to CP violation in flavor-violating
processes, like $\epsilon_K$, and depend on the underlying flavor theory.

Finally we want to point out a major cosmological advantage of the theories
discussed here.
Since there is no light
($\sim m_{3/2}$) gauge singlet field with Planck-scale expectation
value, there is no cause for the
cosmological Polonyi problem~\cite{Coughlan,Roulet,BKN}.

\section{$\mu$ Problem}

An important virtue of hidden sector supersymmetry breaking is the
ease of generating the $\mu$ parameter in the low-energy superpotential at
the correct order of magnitude.  This
mechanism~\cite{giumas} relies on the operator
\begin{equation}
  \int d^4 \theta \frac{z^*}{M_*} H_u H_d,
\end{equation}
where $z$ is a hidden sector field with an $F$-component, $F_z =
{\cal O}(\mu_{SUSY}^2)$.  Then this operator produces
$\mu = {\cal O}(\mu_{SUSY}^2/M_*)$, which is appropriately of the order of the
weak scale.
If there is no gauge-singlet field in the hidden sector, however, the
above operator is forbidden.  Then a natural question is what
alternatives are possible.

%Let us first start with the electroweak symmetry breaking in the
%MSSM.  The potential minimization gives
%\begin{eqnarray}
%        \frac{m_{Z}^{2}}{2} & = & -\mu^{2} +
%                \frac{m_{H_{d}}^{2} - m_{H_{u}}^{2} \tan^{2}\beta}
%                {\tan^{2} \beta -1},
%                \label{eq:mueq}\\
%        2 B\mu &  =  & (2 \mu^{2} + m_{H_{d}}^{2} + m_{H_{u}}^{2})\sin
%        2\beta ,
%        \label{eq:m3eq}
%\end{eqnarray}
%together with the constraints
%\begin{eqnarray}
%  & & m_{H_{d}}^{2} + m_{H_{u}}^{2}  > 0, \\
%  & & m_{H_{d}}^{2} m_{H_{u}}^{2} < (B \mu)^2 .
%\end{eqnarray}
%{\bf Some quantitative comments on the size of $\mu$ and $B\mu$.}

If the hidden sector model is vector-like \cite{IY2,IT1},
the $\mu$ term can be generated by the operator
\begin{equation}
\int d^2 \theta \frac{QQ}{M_*} H_u H_d .
\label{opmu}
\end{equation}
For instance
the fields $Q$ can be chosen to
belong to the $SP(N)$ gauge group in the hidden sector
together with the superpotential $W = \lambda S_{ij} Q^i Q^j$.
If
$N_f = N_c + 1$, the quantum modified constraint $\mbox{Pf} (Q^i Q^j)
= \Lambda^{2 N_f}$ does not allow a supersymmetric vacuum consitent
with the requirement $F_S = 0$.  In a limit where $\lambda$ can be
regarded as perturbative, the quantum modified constraint forces many of
the $Q^i Q^j$ meson operators to acquire expectation values of
${\cal O}(\Lambda^2)$.  The supersymmetry breaking scale is $\mu_{SUSY}^2
\sim \lambda \Lambda^2$.  If $\lambda$ is order unity, the quantum
modified constraint may not be satisfied exactly; one cannot reliably
calculate the meson operator expectation values.  Still, we expect the
meson operators to have expectation values of the same order of
magnitude.  This operator then gives rise to a $\mu$-parameter of
weak-scale size.

The operator in Eq.~(\ref{opmu}) also generates a $B\mu$ term
${\cal O}(\mu_{SUSY} m_{3/2})$, if the $Q$ fields acquire non-vanishing
vacuum expectation
values in their auxiliary components. This may seem a generic feature,
but this is not always the case. If the supersymmetry-breaking sector
possesses an $R$-symmetry unbroken at the vacuum, then the fields $Q$
cannot acquire non-vanishing vacuum expectation values in their
auxiliary components.  In fact, an effective theory analysis suggests
that this is indeed the case \cite{Markusetal}.
A $\mu$ term ${\cal O}(m_{3/2})$ is generated,
but no $B\mu$ terms ${\cal O}(\mu_{SUSY} m_{3/2})$. The $B$ term is originated
from the universal contribution $B=-m_{3/2}$, and hence its phase is
always related to the gaugino mass. The assumed $R$ symmetry is a property
of the supersymmetry-breaking sector in the flat limit. The complete
supergravity theory breaks explicitly the $R$ symmetry, in particular
in the constant term in the superpotential chosen to fine tune the
cosmological constant.
% As a consequence, gaugino masses can be
% generated by supergravity couplings.

%The size of the $B\mu$ term is not clear with this mechanism, however;
%it can be either of ${\cal O}(\mu_{SUSY} m_{3/2})$ or ${\cal
%O}(m_{3/2})^{2}$.  It
%depends on if the meson operator acquires an $F$-component expectation
%value, which is related to another unanswered question if the $U(1)_R$
%symmetry of the model is spontaneously broken.

% For instance, in our previous example, one can exclude
% $\langle S \rangle \gtrsim \Lambda$ based on the renormalization-group
% improved effective potential \cite{ddrg,AM1}.  One cannot study the
% behavior of the potential for $\langle S \rangle \lesssim \Lambda$
% reliably, however.  Since $\langle S \rangle = 0$ is a symmetry
% enhanced point of $U(1)_R$ symmetry, the origin is an extremum of the
% potential.  If it is a minimum, the $U(1)_R$ symmetry is not
% spontaneously broken, and we expect vanishing $F$-component for the
% meson operator.  If it is a maximum, $S$ acquires an ${\cal O}(\Lambda)$
% expectation value  that breaks $U(1)_{R}$ spontaneously, and there is
% probably an $F$-component for the meson operator.  It appears quite
% possible that the $F$-component of the meson operator vanishes, and
% therefore the operator in Eq.~(\ref{opmu}) does not generate a $\mu$
% term.

Notice also that the operator
\begin{equation}
\int d^4 \theta \frac{zz^*}{M_*^2} H_u H_d
\end{equation}
generates a $B\mu$ term, after supersymmetry breaking. This term is
of the correct order of magnitude, the weak scale, and therefore it
is not phenomenologically dangerous. However it spoils the simple
relation between the gaugino mass and the $B$ term and it can introduce
irremovable phases. Nevertheless it is easy to imagine that
Peccei-Quinn-like symmetries of the underlying supergravity theory
forbid the
occurance of this operator.

%  If
%so, the only contribution to the $B\mu$ parameter is from the
%universal $B$-term $B=-m_{3/2}$, and hence its phase is always related
%to the gaugino mass.  In this case there is no CP violation in the
%MSSM Lagrangian except possible phases in the scalar mass squared
%matrices and the usual Kobayashi--Maskawa phase.

If the hidden sector is chiral, we cannot find an operator that
generates the $\mu$-term at the desired order of magnidue.  There are
at least three possibilities to consider in this case.  One is the
generation of the $\mu$-parameter from loop diagrams, the second is
from large expectation values and non-renormalizable interactions, and
the third is the Next-to-Minimal Supersymmetric Standard Model
(NMSSM).

The $\mu$-parameter can be generated by a one-loop diagram of
vector-like fields with a $B$-term \cite{Larry}.  For instance, one can
introduce
vector-like fields with the same (opposite) quantum numbers of left-handed
quarks $Q$ ($\bar{Q}$) and right-handed down-quarks $D$ ($\bar{D}$).
With the superpotential
\begin{equation}
        W = Q D H_{d} + \bar{Q} \bar{D} H_{u} + m_{Q} \bar{Q} Q + m_{D}
        \bar{D} D,
\end{equation}
together with the universal $B$-terms for $m_{Q}$ and $m_{D}$, one
generates both $\mu$ and $B\mu$.  Due to an accidental cancellation
\cite{DGP} (which was later interpreted
in Ref.~\cite{GR}), $m^2_{H_u}$ or $m^2_{H_d}$ are not
generated at the one-loop level.

Another possibility is to employ a field with a flat potential lifted
only by non-renormalizable interactions such that it acquires a large
expectation value.  The global symmetry of the model restricts the
possible terms in the superpotential, which then generates the
$\mu$-term at the desired magnitude.  The first of such example was
given in Ref.~\cite{MSY}, with a global Peccei--Quinn symmetry imposed on
the model, which gives an DSFZ-type axion.  All quark, lepton
superfields carry the PQ charge $+1/2$, while the $H_u$ and $H_d$
$-1$.  The $\mu$-term is forbidden in the superpotential.  The model
has two standard-model singlet fields $P(-1)$ and $Q(n)$ and
right-handed neutrinos $N$.  The charge $n$ is a model-dependent
integer.  The allowed superpotential is then
\begin{equation}
  W = Q d H_d + Q u H_u + L e H_d + L N H_u + P N N +
  \frac{1}{M_*^{n-2}} P^n Q + \frac{1}{M_*^{n-2}} H_u H_d P^{n-2} Q.
\end{equation}
Here, we suppressed all coupling constants and retained only the
dependence on the cutoff-scale $M_*$.  The Yukawa coupling $PNN$
induces a negative mass squared for the $P$ field, which together with
the $|P^{n}|^2/M_*^{2n-4}$ potential from $P^n Q/M_*^{n-2}$ term in
the superpotential generates an expectation value $\langle P \rangle =
{\cal O}(m_{3/2} M_*^{n-2})^{1/(n-1)}$.  The supersymmetry breaking effects
in Eq.~(\ref{eq:universal}) give a term $(n-2) m_{3/2} P^n
Q/M_*^{n-2}$ in the potential, which also foces $Q$ to acquire an
expectation value $\langle Q \rangle = {\cal O}(m_{3/2}
M_*^{n-2})^{1/(n-1)}$.  Then the $\mu$-parameter is automatically of
the desired order of magnitude, $\mu = P^{n-2} Q/M_*^{n-2} =
{\cal O}(m_{3/2})$.  Furthermore with the choice $n = 4$, the
right-handed neutrino mass is in the interesting range for the
atmospheric neutrino and the Peccei--Quinn symmetry breaking scale
(axion decay constant) for the axion CDM.

This type of mechanism is not specific to the Peccei--Quinn
symmetry, but it is desirable to have a symmetry that forbids
Planck-scale $\mu$-term to begin with.  Similar mechanisms were used in
Refs.~\cite{LNS,DNNS}.

The NMSSM is presumably also possible to generate the $\mu$-term at the
weak-scale.  In this scheme, however, it suffers from the possible
tadpole problem for the singlet \cite{PS}, especially in the connection to the
triplet-doublet splitting in grand-unified theories.  Even if the theory is
not grand-unified, it still needs to avoid the gravitational
instability problem \cite{BP}.  If both of the problems are avoided by
an appropriate global symmetry, the NSSM can be a viable option.

\section{Conclusions}
%
% We pointed out a model-independent contribution to the gaugino masses
% proportional to the beta-function from supersymmetry breaking in the
% hidden sector.  They are a consequence of the superconformal anomaly
% multiplet.  The trilinear soft supersymmetry breaking couplings
% ($A$-terms) are generated by the same mechanism.
%
We have shown that there is a completely model-independent contribution
to the gaugino masses from SUSY breaking in the hidden sector whose
origin can be traced to the conformal anomaly.
This contribution to the gaugino mass is given {\it exactly} by
\begin{equation}
m_\lambda = \frac{\beta(g^2)}{2 g^2} m_{3/2},
\end{equation}
where $\beta(g^2)$ is the (1PI) beta function.
Trilinear soft SUSY breaking terms are generated by the same mechanism.

%  This mechanism of
% generating gaugino masses does not require a gauge-singlet field in
% the hidden sector which had been regarded essential in viable models.
% Therefore the cosmological Polonyi problem is not a necessary
% consequence of the hidden sector supersymmetry breaking.

This mechanism opens up the possibility of hidden-sector
models without singlets,
which had been regarded as essential to get gaugino masses of order
$m_{3/2}$.
Models without singlets may be attractive for a variety of reasons,
including simplicity, absence of cosmological problems,
and the absence of instabilities to maintaining the hierarchy.
In models without singlets, then the conformal anomaly contribution gives
the leading contribution to the gaugino mass, predicting gaugino
mass ratios that are very different from the conventional GUT relations.
This is a very general prediction that can be tested if superpartners
are observed in future experiments.

% The model-independent contribution to the gaugino masses, however, are a
% one-loop effect because of the anomaly, and are suppressed.  If there
% is no similar suppression in scalar masses, phenomenology requires the
% scalars to be in multi-TeV range and hence it suffers from a
% fine-tuning problem.  The spectrum of the gaugino masses generated by
% this mechanism does not follow the GUT-relation, even when the theory is
% grand-unified.  This leads to a different phenomenology from the
% conventional one.

One issue that must be addressed in models without singlets is the
fact that the gaugino mass is suppressed by a loop factor compared to
$m_{3/2}$.
Generally, one expects that the scalar masses are of order $m_{3/2}$,
which requires fine-tuning to get acceptable electroweak symmetry
breaking.
This scenario has a phenomenology that is very different from the
conventional one, and we have analyzed some of the main features
in the paper.
%Another possibility is that there is a mechanism that also suppresses the
%scalar masses compared to $m_{3/2}$, such as `no-scale' models with
%Heisenberg symmetry.

{\bf Note added:}
While completing this paper, we received a paper by
L. Randall and R. Sundrum \cite{RS} which also considers
anomalous contributions to the gaugino mass.
These authors also consider an interesting mechanism to suppress
scalar masses compared to $m_{3/2}$.

\section*{Acknowledgements}
We thank Antonio Riotto and Carlos Wagner for useful discussions.
H.M. also thanks Nima Arkani-Hamed and Lawrence J. Hall.
M.A.L. and H.M. thank the CERN theory group for hospitality
during the initial stages of this work.
This work was supported in part by the U.S. Department of
Energy under Contracts DE-AC03-76SF00098, in part by the National
Science Foundation under grants PHY-95-14797 and PHY-98-02551,
and by the Alfred P. Sloan Foundation.

\appendix

\section{Konishi Anomaly Subtlety}

In this appendix, we discuss a subtlety concerning the soft
supersymmetry breaking parameters $A=0$, $B=-m_{3/2}$ discussed in
section \ref{sec:generality}.  The two contributions were given by
$-3\langle W \rangle$ from the $-3|W|^{2}$ term, and $\phi W_{\phi}$
from the cross term in $|\phi^{*}W + W_{\alpha}|^{2}$.  The latter
contributions are actually a consequence of the kinetic operator $\int
d^{4} \theta (m_{3/2} \theta^{2}) \phi^{*} \phi$, and this operator in
general contains the gaugino mass operator as well \cite{Konishi} when
the equation of motion is used to rewrite the auxiliary component $F$
in terms of $F_{\phi} = (\partial W/\partial \phi)^{*}$:
\begin{equation}
        \phi F_{\phi}^* = \frac{\partial W}{\partial \phi} +
        \phi \frac{g^{2}}{16\pi^{2}}T_{F} \lambda \lambda.
\end{equation}
However, this contribution to the gaugino mass from the Konishi
anomaly is absent in a fully regulated theory.  Since it was
necessary to use fully regulated theory in order to understand the
origin of the gaugino mass, the Konishi anomaly effect is always absent
and this concern is a red herring.

The simplest case to see this is when a matter field is accompanied
by the Pauli--Villars regulator.  In this case the regulator field has
the same supersymmetry-breaking effect in the kinetic term that
cancels the Konishi anomaly.

To check the same cancellation in the ${\cal N}=2$ or ${\cal N}=4$
regularization is
somewhat trickier.  For instance with ${\cal N}=4$ regularization, it appears
that the adjoint chiral multiplets produce gaugino mass from the
Konishi anomaly.  However, the gauge multiplet needs a gauge fixing in
a manifestly supersymmetric manner, which requires three
Faddeev--Popov ghost chiral supermultiplets in the adjoint
representation.  Their kinetic terms produce the opposite Konishi
anomaly.  The same cancellation can be checked with the ${\cal N}=2$
regularization as well.

\def\ijmp#1#2#3{{\it Int. Jour. Mod. Phys. }{\bf #1~}(19#2)~#3}
\def\pl#1#2#3{{\it Phys. Lett. }{\bf B#1~}(19#2)~#3}
\def\zp#1#2#3{{\it Z. Phys. }{\bf C#1~}(19#2)~#3}
\def\prl#1#2#3{{\it Phys. Rev. Lett. }{\bf #1~}(19#2)~#3}
\def\rmp#1#2#3{{\it Rev. Mod. Phys. }{\bf #1~}(19#2)~#3}
\def\prep#1#2#3{{\it Phys. Rep. }{\bf #1~}(19#2)~#3}
\def\pr#1#2#3{{\it Phys. Rev. }{\bf D#1~}(19#2)~#3}
\def\np#1#2#3{{\it Nucl. Phys. }{\bf B#1~}(19#2)~#3}
\def\mpl#1#2#3{{\it Mod. Phys. Lett. }{\bf #1~}(19#2)~#3}
\def\arnps#1#2#3{{\it Annu. Rev. Nucl. Part. Sci. }{\bf #1~}(19#2)~#3}
\def\sjnp#1#2#3{{\it Sov. J. Nucl. Phys. }{\bf #1~}(19#2)~#3}
\def\jetp#1#2#3{{\it JETP Lett. }{\bf #1~}(19#2)~#3}
\def\app#1#2#3{{\it Acta Phys. Polon. }{\bf #1~}(19#2)~#3}
\def\nc#1#2#3{{\it Nuovo Cim. }{\bf A#1~}(19#2)~#3}
\def\rnc#1#2#3{{\it Riv. Nuovo Cim. }{\bf #1~}(19#2)~#3}
\def\ap#1#2#3{{\it Ann. Phys. }{\bf #1~}(19#2)~#3}
\def\ptp#1#2#3{{\it Prog. Theor. Phys. }{\bf #1~}(19#2)~#3}

\end{document}